\documentstyle[twoside,fleqn,espcrc2,epsf]{article}
\newcommand{\AmS}{{\protect\the\textfont2
  A\kern-.1667em\lower.5ex\hbox{M}\kern-.125emS}}

\def\lQ{\Lambda_{QCD}}

\newcommand{\be}{\begin{equation}}
\newcommand{\ee}{\end{equation}}
\newcommand{\bea}{\begin{eqnarray}}
\newcommand{\eea}{\end{eqnarray}}
\def\al{\alpha}
\def\als{\alpha_{s}}
\def\siml{{\
    \lower-1.2pt\vbox{\hbox{\rlap{$<$}\lower6pt\vbox{\hbox{$\sim$}}}}\ }} 

\title{NRQCD, Effective Field Theories and Potential Models}

\author{Antonio Pineda\address{Theory Division, CERN, 1211 Geneva 23,
    Switzerland}% 
        \thanks{Marie Curie Fellow, contract No. ERBFMBICT983405.}
        }

\begin{document}

\begin{abstract}
We review some recent developments on non-relativistic effective field
theories for heavy quark-antiquark systems and how they can bring a well founded 
connection between QCD and potential models.
\end{abstract}

\maketitle

The study of the heavy quark--antiquark system is an old topic (see \cite{Yndurain}
for earlier references). Here we will concentrate on recent
developments based on effective field theories. For large enough
masses, these systems can be considered to be non-relativistic (NR) and are
then characterized by, at least, three widely separated scales: hard (the mass
$m$, of the heavy quarks), soft (the relative momentum of the
heavy-quark--antiquark $|{\bf p}| \sim mv$, $ v \ll 1$), and ultrasoft (US, the
typical kinetic energy $E \sim mv^2$ of the heavy quark in the bound state
system). In 1986, NRQED \cite{NRQED}, an effective field theory for
non-relativistic leptons, was presented, providing the first and decisive link 
in a chain of developments that is still growing. NRQED is obtained from QED by
integrating out the hard scale $m$ (see \cite{apNRQED} for some
applications).  It is characterized by an ultraviolet cut-off much smaller
than the mass $m$ and much larger than any other scale. NRQCD \cite{NRQCD1} was
born soon afterwards. NRQCD has proved to be extremely successful in studying
$Q$-${\bar Q}$ systems near threshold. The Lagrangian of NRQCD can be
organized in powers of $1/m$, thus making explicit the non-relativistic
nature of the physical systems.

In order for effective field theories to be useful, a power counting is
needed. The first power counting (organized in powers of $v$ and $\als$) used
to assess the relative importance of the NRQCD matrix elements followed from
arguments valid in the perturbative regime \cite{pc}. In particular, it was
assumed that $\lQ \siml mv^2 $. This implies that dynamics is mainly
perturbative, with $v \sim \als$, dictated by the solution of the
Schr\"odinger equation with the Coulomb potential $-C_f\als/r$. Therefore, it
is debatable whether it could be applied for the charmonium system or for
higher states of the bottomonium systems.  Moreover, even in the perturbative
case, because of the different scales involved in the problem, the matrix
elements do not have a unique power counting but they also contribute to
subleading orders in the $v$ counting, whereas the power counting in Ref.
\cite{pc} only provided the leading order.

At this stage, there were two major questions (somewhat related) in the
formulation of NRQCD: 1) the first was the non-existence of explicit power
counting rules to systematically incorporate subleading effects, even in the
perturbative situation; 2) the second was that the regularization procedure
was based on cut-off regularization. Attempts to perform the matching between
QCD and NRQCD using dimensional regularization had the drawback that the naive
incorporation of the kinetic term in the quark propagator jeopardizes the
power counting rules.

The solution to the latter problem was first given in
Ref. \cite{Manohar}. There, it was argued that the matching between QCD and
NRQCD in dimensional regularization should be performed just as in HQET --namely, the kinetic term must be
treated as a perturbation and then the matching conditions and their
Lagrangian in both theories are the same (in particular the matching
computation in the effective theory is zero)-- and performed for the terms bilinear in the fermions. For the
four-fermion terms the matching along the same lines was worked out in
Refs. \cite{pNRQCD,Match}. The key point in the above derivations was that, in
order to carry out the matching, it is not so important to know the power
counting of each term in the effective theory as to know that the remaining
dynamical scales of the effective theory are much lower than the mass. The
power counting tells us the relative importance between different operators, but
this does not change the value of the matching coefficients. That is, we only
need 
$$
m \gg |{\bf p}|,\; E,\; \lQ\,.
$$
As a matter of fact, this defines NRQCD in dimensional regularization (if it exists) irrespective of the
relative 
size between $\lQ$ and the soft and ultrasoft scale, in particular, even if
$\lQ$ were the next relevant scale. 

The first problem, to obtain a complete power counting, was first studied
within a mainly perturbative framework and triggered two lines of research:  

A) On the one hand, by trying to classify the different momentum regions existing
in a pure perturbative version of NRQCD and/or to reformulate NRQCD in ways
where some of these regions were explicitly displayed by introducing new
fields in the NRQCD Lagrangian. In particular, in Ref. \cite{Labelle}, within
a QED context, the existence of the ultrasoft
region was first made explicit, besides the soft/potential one (of which the author only considered the
leading order contribution somewhat missing the soft region as defined in
\cite{BS98}), and rules were given to estimate their size. Subsequent work
\cite{Luke:1997hj} also tried to get complete power counting rules
for NRQCD, 
but they also missed the soft region.  Within this philosophy, in Ref.
\cite{BS98}, the most complete classification of (perturbative) momentum
regions to date was made by a rigorous diagrammatic study called the {\it threshold
expansion}, including the soft region missed in these works. It was then
realized that the hard region corresponded to the matching between NRQCD and
QCD as described in Refs. \cite{Manohar,Match}. This allowed the computation
of the electromagnetic current
matching coefficient at two-loops \cite{Beneke:1998jm}. Finally, some work was also done on formulating the threshold expansion of NRQCD within an 
effective Lagrangian formalism \cite{Griesshammer}.
  
B) On the other hand, in parallel, a different approach was worked out in
detail in 
\cite{pNRQCD,wl} to deal with the power-counting problem. It tried to answer the
question: How would we like the effective theory for ${\bar Q}$--$Q$ systems near
threshold to be? The first observation was that we did not want to describe
all the degrees of freedom included in NRQCD, but rather only those with US
energy. Therefore, the unwanted degrees of freedom should be integrated out.
Moreover, we wanted to get a closer connection with a Schr\"odinger-like
formulation for these systems (see also \cite{Lepagenuclear}). The idea was to connect NRQCD with potential
models also, eventually, in the non-perturbative regime (more on that
later). Roughly speaking we wanted our effective theory to be something like:
\begin{eqnarray*}
\,\left.
\begin{array}{ll}
&
\displaystyle{\left(i\partial_0-{{\bf p}^2 \over 2m}-V_0(r)\right)\Phi({\bf r})=0}
\\
&
\displaystyle{+{\rm  corrections\; to\; the\; potential}}
\\
&
\displaystyle{+{\rm interaction \;with\; other\; low-}}
\\
&
\displaystyle{\,\,\; {\rm energy\; degrees \;of\; freedom}}
\end{array} \right\} 
{\rm pNRQCD}
\end{eqnarray*}
where $V_0(r)=-C_f\als/r$ in the perturbative case and $\Phi({\bf r})$ is the
${\bar Q}$--$Q$ wave-function.

The resulting effective field theory was called potential NRQCD (pNRQCD).
This is obtained after integrating out the soft scale, understood in the
following way: pNRQCD is defined by two ultraviolet (UV) cut-offs $\Lambda_1$ and
$\Lambda_2$. The former fulfils the relation $ mv^2 \ll \Lambda_1 \ll mv$ and
is the cut-off of the energy of the fermions and of the energy and the
three-momentum of the gluons, whereas the latter fulfils $mv \ll \Lambda_2 \ll
m$ and is the cut-off of the relative three-momentum of the heavy
fermion--antifermion system, ${\bf p}$. This choice of the cutoffs can be
motivated as follows. We are only interested in the degrees of freedom with US
energy and, as a general philosophy, we would like to keep the UV cut-offs of
the effective theory as low as possible. Therefore, any degree of freedom with
larger energy should be integrated out. This fixes $\Lambda_1$ (we also fix
the three-momentum cut-off of the gluons to be $\Lambda_1$ since the gluons
satisfy a relativistic dispersion relation). The motivation for the choice of
$\Lambda_2$ is that fermions with US energy have soft three-momentum owing to
their non-relativistic dispersion relation. In short, we have only integrated
out the degrees of freedom where a perturbative, order by order in $\al$,
matching calculation can be performed (note that heavy fermions with soft
three-momentum and US energy cannot be integrated out perturbatively since
they are the responsible for producing the bound state). We believe that pNRQCD,
defined in this way, has several advantges:

i) The matching between NRQCD and pNRQCD can be done order by order in $\als$
and in $1/m$ (by analytically expanding the NRQCD Feynman diagrams in terms of
the remaining dynamical scales of pNRQCD there is also an expansion in terms of
the ultrasoft scale). In fact, this basically corresponds to obtaining the potential of
the Schr\"odinger-like equation. 

ii) The final effective theory resembles very much a Schr\"odinger-like
equation. We believe that pNRQCD as described above provides a rigorous
conection between NRQCD (quantum field theories) and potential models
(non-relativistic quantum-mechanics formulation) in the situation where $\lQ
\siml mv^2$.

iii) In spite of working with a Schr\"odinger-like equation, US gluons are
still incorporated in a second-quantized, systematic and gauge-invariant
fashion.  

iv) The construction of pNRQCD follows a step-by-step procedure. It goes along
the standard effective field theory idea of integrating out scale by scale
\cite{Georgi}, although properly speaking it should be said region by region.
Since we are dealing with non-relativistic theories, the energy and the
three-momentum are not related by the relativistic dispersion relation but by
the non-relativistic one, producing asymmetric ultraviolet cut-off in the
effective theories. The advantage of working region by region is that in each
(perturbative) region the Feynman integrals become homogeneous\footnote{I
  thank M. Beneke for stressing this point to me.}, and much easier to compute
in dimensional regularization, with just the one scale that we want to
integrate out (on general grounds the matching calculation in each new
effective theory is zero). On the other hand even in a non-perturbative
situation it is expected that a better control in the dynamics would be
obtained. In particular, perturbative regions can be disantangled from
non-perturbative ones in a better controlled way.

Finally we would like to note that, once the UV cut-offs of pNRQCD are fixed,
no ambiguity is left in the definition of pNRQCD. Still, for an easy
comparison with the first line of research mentioned above, we mention that
pNRQCD is obtained from NRQCD after integrating out soft quarks and gluons and
potential gluons. We would like to stress here that, in particular, the
so-called soft region was already included in Ref. \cite{pNRQCD} that, as a
matter of fact, was the first place where a power counting including the
hard, soft, potential and ultrasoft region was given within an effective field
theory framework.

Leaving aside non-perturbative effects, one could be worried about
smaller dynamical scales $mv^3$,... that could appear by fine-tuning the
energy to be near some poles. We would like to note that these eventual
scales will not be a problem since they would be encoded in pNRQCD, but
then it may be convenient to integrate out some further degrees of
freedom to accurately describe physics at these smaller scales.

The developments in effective field theories explained above (with $\lQ \siml
mv^2$) can be applied to several physical situations such as the
$\Upsilon(1S)$ \cite{Pineda:1998hz}, bottomonium sum rules
\cite{Melnikov:1999ug} or $t$--$\bar t$ production near threshold
\cite{Hoang:2000yr}, where we believe that put on more solid theoretical
grounds the formalism used.

There has been some recent attempts to obtain renormalization group equations
for non-relativistic systems. For lack of space we do not discuss them here in
detail (see \cite{RG}). We only briefly comment on the work of \cite{LMR},
from which it could be naively concluded that pNRQCD could not reproduce the
anomalous dimension of the electromagnetic current. This is not true, which
should be evident from the fact that the evaluation of the anomalous dimension
in pNRQCD would go along similar lines to the one performed in
\cite{Melnikov:1999ug}. We believe that the approach in \cite{LMR} may suffer
from the drawback of having to treat the different scales entangled. This is
in contradiction with the philosophy advocated here of working region by
region and may jeopardize the huge simplification obtained in perturbative
calculations. Moreover, since they treat the soft and US scale perturbatively,
it is hard to imagine how to introduce non-perturbative effects in this
approach.

%\medskip

So far we have restricted our considerations to the situation $\lQ \siml
mv^2$. It is doubtful whether we can consider most of the charmonium and
bottomonium spectrum to be in this situation but rather in the (generic)
non-perturbative case with $mv \sim \lQ$. Then, it is not clear, a priori,
what is the power counting that should be used for these systems\footnote{In
  fact, a different, non-standard, power counting of the matrix elements of
  NRQCD may eventually explain the apparent difficulties that NRQCD is facing to explain the polarization of prompt $J/\psi$ data, and to accurately
  determine the different matrix elements (see \cite{Kraemer}).}. In
particular, it is less clear how to obtain a rigorous connection between NRQCD
and potential models (if it exists), although naively one would expect that,
to some extent, the same philosophy as used previously to obtain pNRQCD could
also be followed here. In this line of thinking, we would like to end by briefly
reporting on some new results in our understanding of NRQCD and pNRQCD in
the non-perturbative regime obtained in Ref. \cite{m1}; there, for the first
time, a method to obtain a controlled derivation of the potential from QCD in
terms of Wilson loops at arbitrary orders in $1/m$ has been proposed. This has
permitted to obtain new potentials for the heavy quarkonium (as well as for
the heavy hybrids) at $O(1/m)$ and $O(1/m^2)$ previously missed in the
literature, as well as to correct some of the already existing ones.

%Any potential model for heavy quarkonium should be consistent with our results
%(up to field redefinition). One of the roles of effective field theories: to
%discriminate between different models.

{\bf Acknowledgements}. I thank J. Soto for his careful reading of the manuscript and M. Beneke for
discussions.


\begin{thebibliography}{9}
\bibitem{Yndurain} F.J. Yndur\'ain, ``The Theory of Quarks and Gluon
  Interactions'', third edition, Springer, Heidelberg, 1999.
\bibitem{NRQED} W.E. Caswell and G.P. Lepage, Phys. Lett. {\bf B167}, 437
  (1986).
%\newcommand{\wwwspires}{http://www.slac.stanford.edu/spires/find/hep/www}
%\cite{Kinoshita:1996mt}
\bibitem{apNRQED} T.~Kinoshita and M.~Nio,
%``Radiative Corrections to the Muonium Hyperfine Structure. I. The $\alpha~2 (Z\alpha)$ Correction,''
Phys.\ Rev.\  {\bf D53}, 4909 (1996); 
%%CITATION = HEP-PH 9512327;%%
%\href{\wwwspires?eprint=HEP-PH/9512327}{SPIRES}
%\newcommand{\wwwspires}{http://www.slac.stanford.edu/spires/find/hep/www}
%\cite{Hoang:1997ki}
A.H.~Hoang, P.~Labelle and S.M.~Zebarjad,
%``The single photon annihilation contributions to the positronium  hyperfine splitting to order m(e) alpha**6,''
Phys.\ Rev.\ Lett.\  {\bf 79}, 3387 (1997);
%%CITATION = HEP-PH 9707337;%%
%\href{\wwwspires?eprint=HEP-PH/9707337}{SPIRES}
%\newcommand{\wwwspires}{http://www.slac.stanford.edu/spires/find/hep/www}
%\cite{Czarnecki:1999mw}
A.~Czarnecki, K.~Melnikov and A.~Yelkhovsky,
%``Positronium S state spectrum: Analytic results at O(m alpha**6),''
Phys.\ Rev.\  {\bf A59} 4316  (1999).
%%CITATION = HEP-PH 9901394;%%
%\href{\wwwspires?eprint=HEP-PH/9901394}{SPIRES}


\bibitem{NRQCD1} B.A.~Thacker and G.P.~Lepage, Phys. Rev. {\bf D43}, 196
  (1991). 
\bibitem{pc} G.P. Lepage et al., Phys. Rev. {\bf D46}, 4052 (1992).
%\bibitem{NRQCD} G.T. Bodwin, E. Braaten and G.P. Lepage, Phys. Rev. {\bf D51}, 1125
%  (1995), Erratum {\it ibid.} {\bf D55}, 5853 (1997). 

\bibitem{Manohar} A.V. Manohar, Phys. Rev. {\bf D56}, 230  (1997).
\bibitem{pNRQCD} A. Pineda and J. Soto, Nucl. Phys. {\bf B} (Proc. Suppl.) {\bf 64},
  428 (1998); Phys.\ Lett.\  {\bf B420}, 391 (1998);
%%CITATION = HEP-PH 9711292;%%
%\href{\wwwspires?eprint=HEP-PH/9711292}{SPIRES}
Phys. Rev. {\bf D59}, 016005 (1999).
\bibitem{Match} A. Pineda and J. Soto, Phys. Rev. {\bf D58}, 114011 (1998).
\bibitem{Labelle} P. Labelle, Phys. Rev. {\bf D58}, 093013 (1998). 

\bibitem{BS98} M. Beneke and V.A. Smirnov, Nucl. Phys. {\bf B522}, 321
  (1998). 


%\newcommand{\wwwspires}{http://www.slac.stanford.edu/spires/find/hep/www}
%\cite{Luke:1997hj}
\bibitem{Luke:1997hj}
M.~Luke and A.V.~Manohar,
%``Bound states and power counting in effective field theories,''
Phys.\ Rev.\  {\bf D55}, 4129 (1997);
%%CITATION = HEP-PH 9610534;%%
%\href{\wwwspires?eprint=HEP-PH/9610534}{SPIRES}
%\newcommand{\wwwspires}{http://www.slac.stanford.edu/spires/find/hep/www}
%\cite{Grinstein:1998gv}
B.~Grinstein and I.Z.~Rothstein,
%``Effective field theory and matching in non-relativistic gauge theories,''
Phys.\ Rev.\  {\bf D57}, 78 (1998);
%%CITATION = HEP-PH 9703298;%%
%\href{\wwwspires?eprint=HEP-PH/9703298}{SPIRES}
%\newcommand{\wwwspires}{http://www.slac.stanford.edu/spires/find/hep/www}
%\cite{Luke:1998ys}
M.~Luke and M.J.~Savage,
%``Power counting in dimensionally regularized NRQCD,''
Phys.\ Rev.\  {\bf D57}, 413 (1998).
%%CITATION = HEP-PH 9707313;%%
%\href{\wwwspires?eprint=HEP-PH/9707313}{SPIRES}


%\newcommand{\wwwspires}{http://www.slac.stanford.edu/spires/find/hep/www}
%\cite{Beneke:1998jm}
\bibitem{Beneke:1998jm}
M.~Beneke, A.~Signer and V.A.~Smirnov,
%``Two-loop correction to the leptonic decay of quarkonium,''
Phys.\ Rev.\ Lett.\  {\bf 80}, 2535 (1998);
%%CITATION = HEP-PH 9712302;%%
%\href{\wwwspires?eprint=HEP-PH/9712302}{SPIRES}
%\newcommand{\wwwspires}{http://www.slac.stanford.edu/spires/find/hep/www}
%\cite{Czarnecki:1998vz}
A.~Czarnecki and K.~Melnikov,
%``Two-loop QCD corrections to the heavy quark pair production cross  section in e+ e- annihilation near the
Phys.\ Rev.\ Lett.\  {\bf 80}, 2531 (1998).
%%CITATION = HEP-PH 9712222;%%
%\href{\wwwspires?eprint=HEP-PH/9712222}{SPIRES}

\bibitem{Griesshammer} H.W.~Griesshammer, Phys. Rev. {\bf D 58}, 094027 (1998)

%\bibitem{pNRQED}
%A.~Pineda and J.~Soto,
%%``The Lamb shift in dimensional regularisation,''
%Phys.\ Lett.\  {\bf B420}, 391 (1998);
%%%CITATION = HEP-PH 9711292;%%
%%\href{\wwwspires?eprint=HEP-PH/9711292}{SPIRES}
%Phys. Rev. {\bf D59}, 016005 (1999).

\bibitem{wl} 
N.~Brambilla, A.~Pineda, J.~Soto and A.~Vairo,
Phys.\ Rev.\  {\bf D60}, 091502 (1999); 
Nucl.\ Phys.\  {\bf B566}, 275 (2000).

\bibitem{Lepagenuclear} G.P. Lepage, nucl-th/9706029.

%\newcommand{\wwwspires}{http://www.slac.stanford.edu/spires/find/hep/www}
%\cite{Georgi:1993qn}
\bibitem{Georgi}
H.~Georgi,
%``Effective field theory,''
Ann.\ Rev.\ Nucl.\ Part.\ Sci.\  {\bf 43}, 209 (1993).
%%CITATION = ARNUA,43,209;%%
%\href{\wwwspires?j=ARNUA\%2c43\%2c209}{SPIRES}
%\newcommand{\wwwspires}{http://www.slac.stanford.edu/spires/find/hep/www}
%\cite{Pineda:1998hz}
\bibitem{Pineda:1998hz}
A.~Pineda and F.J.~Yndurain,
%``Calculation of quarkonium spectrum and m(b), m(c) to order  alpha(s)**4,''
Phys.\ Rev.\  {\bf D58}, 094022 (1998).
%%CITATION = HEP-PH 9711287;%%
%\href{\wwwspires?eprint=HEP-PH/9711287}{SPIRES}

%\newcommand{\wwwspires}{http://www.slac.stanford.edu/spires/find/hep/www}
%\cite{Melnikov:1999ug}
\bibitem{Melnikov:1999ug}
K.~Melnikov and A.~Yelkhovsky,
%``The b quark low-scale running mass from Upsilon sum rules,''
Phys.\ Rev.\  {\bf D59}, 114009 (1999);
%%CITATION = HEP-PH 9805270;%%
%\href{\wwwspires?eprint=HEP-PH/9805270}{SPIRES}
A.A.~Penin and A.A.~Pivovarov,
%``Next-to-next-to-leading order vacuum polarization function of heavy  quark near threshold and sum rules for b anti-b
Phys.\ Lett.\  {\bf B435}, 413 (1998); 
Nucl.\ Phys.\  {\bf B549}, 217 (1999); 
A.H.~Hoang,
%``1S and MS-bar bottom quark masses from Upsilon sum rules,''
Phys.\ Rev.\  {\bf D61}, 034005 (2000); 
M.~Beneke and A.~Signer,
%``The bottom MS-bar quark mass from sum rules at next-to-next-to-leading  order,''
Phys.\ Lett.\  {\bf B471}, 233 (1999).
%%CITATION = HEP-PH 9906475;%%
%\href{\wwwspires?eprint=HEP-PH/9906475}{SPIRES}



%\newcommand{\wwwspires}{http://www.slac.stanford.edu/spires/find/hep/www}
%\cite{Hoang:2000yr}
\bibitem{Hoang:2000yr}
A.~H.~Hoang {\it et al.},
%``Top-antitop pair production close to threshold: Synopsis of recent NNLO  results,''
hep-ph/0001286.
%%CITATION = HEP-PH 0001286;%%
%\href{\wwwspires?eprint=HEP-PH/0001286}{SPIRES}


%\bibitem{KP} B.A.~Kniehl and A.A.~Penin, hep-ph/9907489.
\bibitem{RG} A.V.~Manohar, BEACH00 proceedings.
\bibitem{LMR} M.E. Luke, A.V. Manohar and I.Z. Rothstein, Phys. Rev. {\bf D61},
  074025 (2000). 
\bibitem{Kraemer} M. Kraemer, hep-ph/0010137; A.K. Leibovich, hep-ph/0008236.
\bibitem{m1} N. Brambilla, A. Pineda, J. Soto and A. Vairo, hep-ph/0002250;
  A. Pineda and A. Vairo, hep-ph/0009145.
\end{thebibliography}
\end{document}